\documentclass[twocolumn,prl,showpacs]{revtex4}%
\usepackage{graphicx}%
\usepackage{amsmath}%
\usepackage{amsfonts}%
\usepackage{amssymb}
\usepackage{bm}

\def\k{{ {\bm k} }}

\def\q{{ {\bm q} }}
\def\Q{{ {\bm Q} }}

\def\w{{\omega}}
\def\a{{\alpha}}

\begin{document}
\title{
Self-Consistent Vertex Correction Analysis
for Iron-Based Superconductors: \\
Mechanism of Coulomb-Interaction-Driven Orbital Fluctuations
}
\author{Seiichiro \textsc{Onari}$^{1}$ and
Hiroshi \textsc{Kontani}$^{2}$
}
\date{\today }

\begin{abstract}
We study the mechanism of orbital/spin fluctuations
due to multiorbital Coulomb interaction in iron-based superconductors,
going beyond the random-phase-approximation.
For this purpose,
we develop a self-consistent vertex correction (SC-VC) method,
and find that multiple orbital fluctuations 
in addition to spin fluctuations are mutually emphasized by 
the ``multimode interference effect'' described by the VC.
Then, both the antiferro-orbital 
and ferro-orbital (=nematic) fluctuations simultaneously 
develop for $J/U \sim 0.1$, 
both of which contribute to the $s$-wave superconductivity.
Especially, the ferro-orbital fluctuations give the 
orthorhombic structure transition as well as the 
softening of shear modulus $C_{66}$.
\end{abstract}

\address{
$^1$ Department of Applied Physics, Nagoya University and JST, TRIP, 
Furo-cho, Nagoya 464-8602, Japan. \\
$^2$ Department of Physics, Nagoya University and JST, TRIP, 
Furo-cho, Nagoya 464-8602, Japan. 
}
 

\sloppy

\maketitle

%%%%%%%%%%%%%%%%%%
%Introduction
%%%%%%%%%%%%%%%%%%
Since the discovery of iron-based superconductors,
the mechanism of high-$T_{\rm c}$ superconductivity
has been studied very actively.
Theoretically, 
both the spin-fluctuation-mediated $s_\pm$-wave state
(with sign reversal of the gap between hole-pocket (h-pocket) 
and electron-pocket (e-pocket))
\cite{Kuroki,Mazin,Hirschfeld,Chubukov,DHLee}
and the orbital-fluctuation-mediated $s_{++}$-wave state
(without sign reversal)
\cite{Kontani-RPA,Yanagi}
had been proposed.
The latter scenario is supported by the robustness of 
$T_{\rm c}$ against impurities in many iron-pnictides
\cite{Sato-imp,Nakajima,Li,Paglione,Onari-impurity}.
% ARPES,neutron 
Possibility of impurity-induced crossover from $s_\pm$ to $s_{++}$ states
had been discussed theoretically \cite{Kontani-RPA,Hirschfeld}.
Also, orbital-independent gap observed in BaFe$_2$(As,P)$_2$ 
and (K,Ba)Fe$_2$As$_2$ by laser ARPES measurement 
\cite{Shimo-Science,Saito-RPA}
as well as the ``resonance-like'' hump structure 
in the neutron inelastic scattering \cite{Onari-resonance}
are consistent with the orbital fluctuation scenario.
%In addition, experimental ``resonance-like'' hump structure 
%in the neutron inelastic scattering is well reproduced in terms 
%of the $s_{++}$-wave state
%\cite{Onari-resonance,Onari-resonance2}.

Nature of orbital fluctuations has been studied intensively 
after the discovery of large softening of the shear modulus $C_{66}$
\cite{Fer,Goto,Yoshizawa} and 
renormalization of phonon velocity
\cite{neutron-phonon}
observed well above the orthorhombic structure transition temperature $T_S$.
Consistently, a sizable orbital polarization is observed in 
the orthorhombic phase 
\cite{ARPES,Xray}.
Moreover, the ``electronic nematic state'' 
%in the tetragonal phase 
with large in-plane anisotropy of resistivity or magnetization
well above $T_S$ and $T_{\rm c}$
\cite{Fisher,Matsuda},
also indicates the occurrence of (impurity-induced local) 
orbital order \cite{Inoue-Nematic}.

\begin{figure}[!htb]
\includegraphics[width=.99\linewidth]{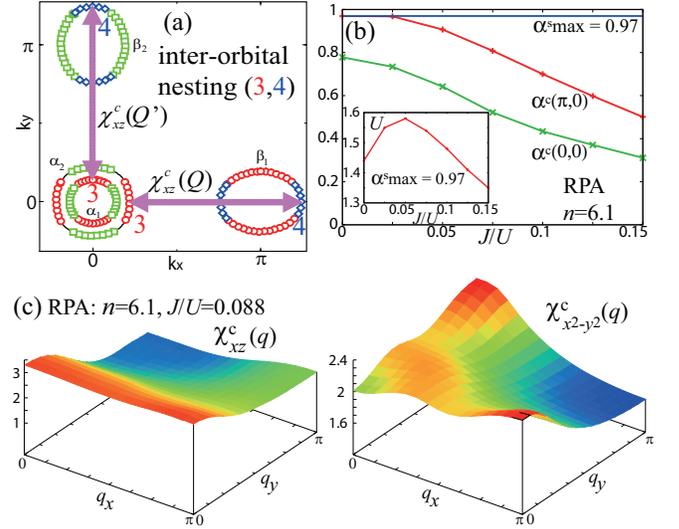}
\caption{(color online)
(a) Fermi surfaces of iron pnictides.
The colors correspond to $2=xz$ (green), $3=yz$ (red), 
and $4=xy$ (blue), respectively.
(b) $\a^{c}_\Q$, $\a^{c}_{\bm 0}$ and $U$
as function of $J/U$ in RPA under the condition $\a^{s}_{\rm max}=0.97$.
(c) $\chi^c_{xz}(\q)$ and $\chi^c_{x^2-y^2}(\q)$
%and (d) $\chi^s(\q)={\rm Tr}\{ {\hat\chi}^s(\q) \}$
given by the RPA for $(J/U,U)=(0.088,1.53)$.
%$\chi^c_{xz}$ is weakly enlarged by the inter-orbital ($3,4$) nesting.
}
\label{fig:chiRPA}
\end{figure}

Origin of orbital order/fluctuation
had been actively discussed, 
mainly based on the multiorbital Hubbard model with
intra (inter) orbital interaction $U$ ($U'$) and 
the exchange interaction $J=(U-U')/2>0$
\cite{PP,Kontani-RPA}.
We had focused attention to a good {\it inter-orbital nesting}
of the Fermi surfaces shown in Fig. \ref{fig:chiRPA} (a): 
Although moderate orbital fluctuations are induced by $U'$
in the random-phase-approximation (RPA), 
the spin susceptibility due to the intra-orbital nesting, $\chi^s(\q)$, 
is the most divergent for $J>0$ ({\it i.e.}, $U>U'$).  
Since ${J}/{U}\approx 0.12-0.15$ 
according to the first-principle study \cite{Miyake-Arita},
the RPA fails to explain experimental ``nonmagnetic'' structure transition.
This situation is unchanged even if the self-energy correction 
is considered in the fluctuation-exchange (FLEX) approximation
 \cite{Onari-FLEX}.

To explain the strong development of orbital fluctuations,
we had introduced a quadrupole interaction \cite{Kontani-RPA}:
\begin{eqnarray}
H_{\rm quad}=-g\sum_{i} \left( 
{\hat O}_{xz}^i{\hat O}_{xz}^i+{\hat O}_{yz}^i{\hat O}_{yz}^i \right)
 \label{eqn:Hquad}
\end{eqnarray}
where $g$ is the coupling constant, and ${\hat O}_{\gamma}$ is 
the charge quadrupole operator ;$\gamma=xz,yz,xy,x^2-y^2,3z^2-r^2$.
(Hereafter, $x,y$-axes ($X,Y$-axes) are along the nearest Fe-Fe 
(Fe-As) direction.)
This term is actually caused by the electron-phonon ($e$-ph) 
coupling due to in-plane Fe-ion oscillations
 \cite{Kontani-RPA,Onari-FLEX,Saito-RPA}.
Since ${\hat O}_{xz(yz)}$ induces the inter-orbital scattering,
strong antiferro (AF) orbital fluctuations develop for $g\gtrsim0.2$eV
owing to a good inter-orbital nesting.
We also studied the vertex correction (VC)
beyond the RPA \cite{Kontani-Soft},
and obtained strong enhancement of ferro-quadrupole 
(${\hat O}_{x^2-y^2}\propto {\hat n}_{xz}-{\hat n}_{yz}$) 
susceptibility $\chi^c_{x^2-y^2}({\bm 0})$,
which causes the orthorhombic structure transition 
and the softening of $C_{66}$ \cite{Kontani-Soft}.
This ``nematic fluctuation'' is derived from
the interference of two AF orbitons due to the symmetry relation 
${\hat O}_{x^2-y^2}({\bm 0}) \sim {\hat O}_{XZ}(\Q)\times {\hat O}_{YZ}(-\Q)$,
where ${\hat O}_{XZ(YZ)}=[{\hat O}_{xz}+(-){\hat O}_{yz}]/\sqrt{2}$.
%Physically, ferro-$O_{x^2-y^2}$ order is induced 
%by the bound-state formation of two AF orbitons.
%aaa
Then, it was natural to expect that
such multi-orbiton interference effect, which is
given by the VC while dropped in the RPA,
induces large ``Coulomb-interaction-driven'' orbital fluctuations.

In this letter, we study the orbital and spin fluctuations
in iron-based superconductors
by considering the multiorbital Coulomb interaction
with $U=U'+2J$ and $J/U\sim O(0.1)$.
%To take the multi-orbiton(magnon) process into account beyond the RPA,
We develop the self-consistent-VC (SC-VC) method, 
and find that both ferro-$O_{x^2-y^2}$ and AF-$O_{xz/yz}$ fluctuations
strongly develop even for $J/U\sim 0.1$,
due to the inter-orbital nesting and the 
positive interference between multi-fluctuation
(orbiton+magnon) modes.
This result leads to a conclusion that RPA {\it underestimates} 
the orbital fluctuations in multiorbital systems.
%due to the mutual positive feedback.
The present study offers a unified explanation for 
both the superconductivity and structure transition
in many compounds.

%%%%%%%%%%%%%%%%%%%%%%

%susceptibility 
Here, we study the five-orbital Hubbard model
introduced in Ref. \cite{Kuroki}.
We denote $d$-orbitals $m=3z^2-r^2$, $xz$, $yz$, $xy$, and $x^2-y^2$
as 1, 2, 3, 4 and 5, respectively.
The Fermi surfaces are mainly composed of orbitals 2, 3 and 4 
\cite{Kontani-Soft}.
Then, the susceptibility for the charge (spin)
channel is given by the following $25\times25$ matrix form 
in the orbital basis:
\begin{eqnarray}
{\hat \chi}^{c(s)}(q)={\hat {\chi}}^{{\rm irr},c(s)}(q)
 (1-{\hat \Gamma}^{c(s)}{\hat {\chi}}^{{\rm irr},c(s)}(q))^{-1} ,
\label{eqn:chi}
\end{eqnarray}
where $q=(\q,\w_l=2\pi l T)$, and
${\hat \Gamma}^{c(s)}$ represents the Coulomb interaction
for the charge (spin) channel 
composed of $U$, $U'$ and $J$ given in Refs. \cite{Kontani-RPA,Saito-RPA}.
The irreducible susceptibility in Eq. (\ref{eqn:chi}) is given as
\begin{eqnarray}
{\hat {\chi}}^{{\rm irr},c(s)}(q)= {\hat \chi}^{0}(q)+{\hat X}^{c(s)}(q) ,
\label{eqn:irr}
\end{eqnarray}
where $\chi^{0}_{ll',mm'}(q)=-T\sum_p G_{lm}(p+q)G_{m'l'}(p)$ is the bare bubble,
and the second term
%$\Delta{\hat {\chi}}^{{\rm irr},c(s)}$ 
is the VC (or orbiton or magnon self-energy) 
that is neglected in both RPA and FLEX approximation.
In the present discussion, it is convenient to consider the 
quadrupole susceptibilities:
\begin{eqnarray}
\chi^c_{\gamma,\gamma'}(q)&\equiv&
\sum_{ll',mm'} O_\gamma^{l,l'}{\chi}_{ll',mm'}^c(q)O_{\gamma'}^{m',m}
\nonumber \\
&=&{\rm Tr}\{ {\hat O}_\gamma {\hat \chi}^c(q){\hat O}_{\gamma'} \} .
%\chi^Q_{\Gamma\Gamma'}(q)
%\equiv{\rm Tr}\{{\hat O}_\Gamma {\hat \chi}^c(q){\hat O}_{\Gamma'}\}
\label{eqn:chiQ}
\end{eqnarray}
%
%which is expressed as 
%${\rm Tr}\{ {\hat O}_\gamma {\hat \chi}^c(q){\hat O}_{\gamma'} \}$.
Non-zero matrix elements of the quadrupole operators 
for the orbital $2\sim4$ are
$O_{xz}^{3,4}=O_{yz}^{2,4}=O_{x^2-y^2}^{2,2}=-O_{x^2-y^2}^{3,3}=1$
 \cite{Kontani-Soft}. 
%and $l_{x}^{2,4}=l_{y}^{4,3}=l_{z}^{3,2}=i$.
Because of the symmetry, the off-diagonal susceptibilities
($\gamma\ne\gamma'$)
are zero or very small for $\q={\bm 0}$ and the nesting vector
$\Q\approx (\pi,0)$ or $\Q'\approx (0,\pi)$ \cite{Kontani-Soft}.
We do not discuss the angular momentum (dipole) susceptibility,
$\chi_\mu^c(\q)\sim \langle {\hat l}_\mu(\q){\hat l}_\mu(-\q)\rangle$,
since it is found to be suppressed by the VC.
Note that 
${\hat O}_{\mu\nu}\propto {\hat l}_\mu {\hat l}_\nu 
+{\hat l}_\nu {\hat l}_\mu$.

To measure the distance from the criticality,
we introduce the charge (spin) Stoner factor $\a^{c(s)}_\q$, which is 
the largest eigenvalue of 
${\hat \Gamma}^{c(s)}{\hat {\chi}}^{{\rm irr},c(s)}(\q)$ at $\w_l=0$:
The charge (spin) susceptibility diverges when
$\a^{c(s)}_{\rm max}\equiv {\rm max}_\q\{\a^{c(s)}_\q\}=1$.
In a special case $J=0$, the relation $\a^s_{\rm max}=\a^c_{\rm max}$
holds at the momentum $\Q$ in the RPA;
see Fig. \ref{fig:chiRPA} (b).
That is, both spin and orbital susceptibilities 
are equally enhanced at $J=0$, which is unchanged by the self-energy 
correction in the FLEX approximation \cite{Onari-FLEX}.
For $J>0$, the spin fluctuations are 
always dominant ($\a^s_{\rm max}>\a^c_{\rm max}$) in the RPA or FLEX.
However, because of large ${\hat X}^c(q)$,
the opposite relation $\a^s_{\rm max}\lesssim\a^c_{\rm max}$ can be realized 
even for $J/U\lesssim 0.1$ in the SC-VC method.

%%%%%%%%%%%%%%%%%%%%%%%%%%%%%%%%%
%%%%%%%%%%%%%%%%%%%%%%%%%%%%%%%%%

%RPA result
First, we perform the RPA calculation for $n=6.1$ and $T=0.05$,
using $32\times32$ $\k$-meshes:
The unit of energy is eV hereafter.
Figure \ref{fig:chiRPA} (c) shows the diagonal 
quadrupole susceptibilities for $J/U=0.088$;
$\chi^c_{\gamma}(q)\equiv \chi^c_{\gamma\gamma}(q)$.
(The spin susceptibility is shown in Ref. \cite{Kuroki}.)
The Stoner factors are 
$\a^s_{\rm max}=0.97$, $\a^c_\Q=0.76$, and $\a^c_{\bm 0}=0.47$;
see Fig. \ref{fig:chiRPA} (b).
In the RPA, $\chi^c_{xz}(\Q)$ [$\chi^c_{yz}(\Q')$] 
is weakly enlarged by the inter-orbital 
($3,4$) [($2,4$)] nesting,
while $\chi^c_{x^2-y^2}(\q)$ is relatively small and AF-like.
Thus, the RPA cannot explain the structure transition
that requires the divergence of $\chi^c_{x^2-y^2}({\bm 0})$.

\begin{figure}[!htb]
\includegraphics[width=.99\linewidth]{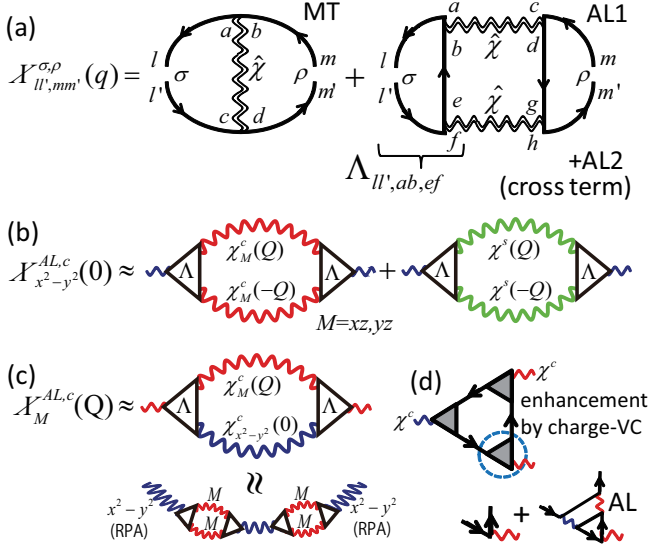}
\caption{(color online)
(a) 
The MT and AL terms:
The wavy and solid lines are susceptibilities
and electron Green functions, respectively.
$\Lambda_{ll',ab,ef}$ is the three-point vertex.
(b) Dominant AL terms for $\chi^c_{x^2-y^2}({\bm 0})$;
the first (second) term represents the two-orbiton (two-magnon) process.
(c) Dominant AL terms for $\chi^c_{M}(\Q)$ ($M=xz,yz$);
higher-order terms with bubbles made of $\chi^c_{M}(\pm\Q)$ 
(= multi-fluctuation process) are relevant.
(d) Enhancement of $\Lambda_{ll',ab,ef}$ due to charge VCs.
}
\label{fig:VC}
\end{figure}

Next, we study the role of VC
due to the Maki-Thompson (MT) and Aslamazov-Larkin (AL) terms
in Fig. \ref{fig:VC} (a),
which become important near the critical point
\cite{Moriya,Bickers}.
Here, ${\hat X}^{c(s)}(q)\equiv {\hat X}^{\uparrow,\uparrow}(q) +(-)
{\hat X}^{\uparrow,\downarrow}(q)$, and wavy lines represent $\chi^{s,c}$.
The AL term (AL1+AL2) for the charge sector,
$X_{ll',mm'}^{{\rm AL},c}(q)$, is given as
\begin{eqnarray}
& &\frac{T}2\sum_{k}\sum_{a\sim h}
\Lambda_{ll',ab,ef}(q;k)\{  {V}_{ab,cd}^c(k+q){V}_{ef,gh}^c(-k)
\nonumber \\
& &\ \ +3{V}_{ab,cd}^s(k+q){V}_{ef,gh}^s(-k) \}
\Lambda_{mm',cd,gh}'(q;k) ,
 \label{eqn:ALexample}
\end{eqnarray}
where 
${\hat V}^{s,c}(q)\equiv{\hat \Gamma}^{s,c}
+ {\hat\Gamma}^{s,c}{\hat\chi}^{s,c}(q){\hat\Gamma}^{s,c} $, 
${\hat \Lambda}(q;k)$ is the three-point vertex
made of three Green functions in Fig. \ref{fig:VC} (a) \cite{Kontani-Soft}, and 
$\Lambda_{mm',cd,gh}'(q;k)\equiv
\Lambda_{ch,mg,dm'}(q;k)+\Lambda_{gd,mc,hm'}(q;-k-q)$.
We include all $U^2$-terms, which are important for reliable results.
The expressions of other VCs will be published in future.

%%%%%%%%%%%%%%%%%%%%%%%%%%%%%%%%%
%%%%%%%%%%%%%%%%%%%%%%%%%%%%%%%%%

Both MT and AL terms correspond to the first-order
mode-coupling corrections to the RPA susceptibility:
The intra- (inter-) bubble correction gives the MT (AL) term
\cite{Moriya}.
%They are also given by the 
%Ward identity for the FLEX self-energy: 
%${\hat X}=\delta{\hat \Sigma_{\rm FLEX}}/\delta{\hat G}$.
In single-orbital models, the VC due to MT+AL terms
had been studied by the self-consistent-renormalization (SCR)
theory \cite{Moriya} or
FLEX approximation with VC \cite{Bickers},
and successful results had been obtained.
In the former (latter) theory, the susceptibility is calculated
in the self-consistent (self-inconsistent) way.
%Here, we present the first study of the MT+AL terms
%in the multiorbital Hubbard model, 
Here, we find a significant role of the AL term 
inherent in the multiorbital Hubbard model.

\begin{figure}[!htb]
\includegraphics[width=.99\linewidth]{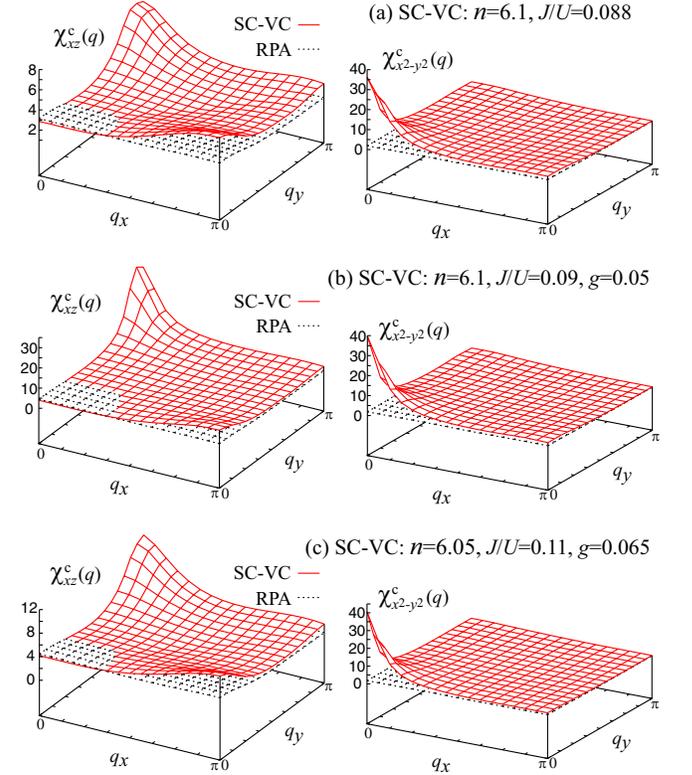}
\caption{(color online)
$\chi^c_{xz}(\q)$ and $\chi^c_{x^2-y^2}(\q)$ given by the SC-VC method.
The relation $\a^s_{\rm max}=\a^c_{\bm 0}=0.97$ is satisfied in all cases:
(a) $n=6.1$ and $J/U=0.88$ ($\a^c_\Q=0.86$),
(b) $n=6.1$, $J/U=0.9$ and $g=0.05$ ($\a^c_{\Q}=0.96$), and
(c) $n=6.05$, $J/U=0.11$ and $g=0.065$ ($\a^c_{\Q}=0.87$).
%The cutoff energy of the quadrupole interaction is
%$\w_D=4T=0.2$ \cite{Kontani-RPA}.
}
\label{fig:chiQ}
\end{figure}

Now, we perform the SC-VC analysis, in the way to satisfy 
${\hat \chi}^{c,s}(q)$ in the VC are equal to the 
total susceptibilities in Eq. (\ref{eqn:chi}).
Then, ${\hat \chi}^{c}(q)$ is strongly enhanced
by $X^{{\rm AL},c}$ in Eq. (\ref{eqn:ALexample}), 
which is relevant when either ${\hat \chi}^c$ or ${\hat \chi}^s$ is large.
On the other hand, we have verified numerically that 
${\hat X}^{s}\sim T\sum \Lambda\cdot V^sV^c \cdot \Lambda$ is 
less important, although it could be relevant only when 
both ${\hat \chi}^c$ and ${\hat \chi}^s$ are large.
Hereafter, we drop ${\hat X}^{s}(q)$ to simplify the argument.
%%%
Figure \ref{fig:chiQ} (a) show $\chi^c_{\gamma}(\q)$ 
given by the SC-VC method for $n=6.1$, $J/U=0.088$ and $U=1.53$,
in which the Stoner factors are
$\a^s_{\rm max}=\a^c_{\bm 0}=0.97$ and $\a^c_\Q=0.86$.
%%%
Compared to the RPA,
both $\chi^c_{x^2-y^2}(\q)$ and $\chi^c_{xz}(\q)$ are strongly enhanced 
by the charge AL term, ${\hat X}^{{\rm AL},c}$, since the  
results are essentially unchanged even if MT term is dropped.
%(On the other hand, unphysical results such as
%negative susceptibilities are frequently obtained if AL term is dropped.)
%The relation $\a^s_{\rm max}<\a^c_{\rm max}=0.97$ is realized
%for $J/U<(J/U)_c=0.088$.
%, and it increases to 0.095 for $n=6.0$.
In the SC-VC method, the enhancements of other 
charge multipole susceptibilities are small.
Especially, both the density and dipole susceptibilities,
$\sum_{l,m}{\hat \chi}^c_{ll,mm}(\q)$ and $\chi^c_\mu(q)$ ($\mu=x,y,z$)
respectively, are suppressed.

Here, we discuss the importance of the AL term:
At $\q\approx {\bm 0}$ or $\Q$, $\chi^c_\gamma(\q)$ 
is enlarged by the diagonal vertex correction with respect to $\gamma$,
$X^{{\rm AL},c}_\gamma(\q) \equiv {\rm Tr}\{ {\hat O}_\gamma 
{\hat X}^{{\rm AL},c}(\q){\hat O}_\gamma \}/{\rm Tr}\{{\hat O}_\gamma^2\}$,
since the off-diagonal terms are absent or small \cite{Kontani-Soft}.
The charge AL term in Eq. (\ref{eqn:ALexample})
is given by the products of two $\chi^c$'s (two-orbiton process)
and two $\chi^s$'s (two-magnon process),
shown in Fig. \ref{fig:VC} (b).
The former process was discussed in Ref. \cite{Kontani-Soft},
and the latter has a similarity to the spin nematic 
theory in Ref. \cite{Fer} based on a frustrated spin model.
Now, we consider the orbital selection rule for the two-orbiton process:
%due to ${\hat \Lambda}$:
Because of the relation 
${\rm Tr}\{{\hat O}_{x^2-y^2} {\hat O}_{M}^2\}\ne0$ for $M=xz,yz$
and a rough relation
$\Lambda_{ll',ab,cd}\sim \Lambda_{ll',l'b,bl}\delta_{l',a}\delta_{b,c}\delta_{d,l}$
 \cite{Kontani-Soft},
the two-orbiton process for $\gamma=x^2-y^2$
is mainly given by $\chi_{M}^c(\Q)^2$.
According to Eq. (\ref{eqn:ALexample}) and Ref. \cite{Kontani-Soft},
$X^{{\rm AL},c}_{x^2-y^2}({\bm 0})\sim
\Lambda^2 U^4 T\sum_q \{\chi(q)\}^2$
grows in proportion to $T\chi(\Q)$ 
[$\log \{\chi(\Q)\}^2$] at high [low] temperatures.
%The analytic expression for the two-orbiton exchange process
%is represented in Ref. \cite{Kontani-Soft}.
In the case of Fig. \ref{fig:chiQ} (a),
two-magnon process is more important for $\chi^c_{x^2-y^2}({\bm 0})$
because of the relation $\a^s_\Q>\a^c_\Q$.
We checked that the two-magnon process is mainly caused by 
$\chi^s_{22,22}(\Q)^2-\chi^s_{22,33}(\Q)^2>0$.

%$\chi^s_{22,22}(\Q)$ and $\chi^s_{33,33}(\Q)$;
%note that $\chi^s_{22,22}(\Q)>\chi^s_{22,33}(\Q)$.
%the octuple fluctuation,
%$\chi^s_{M}(\Q)\equiv {\rm Tr}\{ {\hat O}_{M}{\hat \chi}^s(\Q)
%{\hat O}_{M} \}$ ($M=xz,yz$),
%given by the combination of spin and orbital degrees of freedoms.

%%%%%%%%%%%%%%%%%%%%%%%%%%%%%%%%%
%%%%%%%%%%%%%%%%%%%%%%%%%%%%%%%%%

In the same way, $X^{c}_M(\Q)
\sim\Lambda^2 U^4 T\sum_q \chi^c_{M}(q+Q)\chi^c_{x^2-y^2}(q)$
is enlarged by the two-orbiton process 
due to $\chi^{c}_M(\Q)$ and $\chi^c_{x^2-y^2}({\bm 0})$,
shown in Fig. \ref{fig:VC} (c).
(In this case, two-magnon process is less important
since $\chi^s({\bm 0})$ is small.)
%$X^{{\rm AL},c}_{M}(\Q)\sim
%\Lambda^2 U^4 T\sum_q \chi^c_{M}(q+Q)\chi^c_{x^2-y^2}(q)$.
The obtained $\chi^c_{xz}(\q)$ has peaks at $\q=\Q$ and $\Q'$
since the inter-orbital scattering is emphasized by 
$X^{c}_{xz}(\Q)\propto \chi^c_{x^2-y^2}({\bm 0})\gg1$.
Thus, both $\chi^c_{xz}(\Q)$ and $\chi^c_{x^2-y^2}({\bm 0})$
are strongly enlarged in the SC-VC method,
because of the ``positive feedback'' brought by these two AL terms:
Figure \ref{fig:VC} (c) shows an example of the higher-order terms
that are automatically generated in the SC-VC method.
Such ``multi-fluctuation processes''
inherent in the self-consistent method magnify the RPA results.

Thus, strong ferro- and AF-orbital fluctuations are caused by AL terms.
Both fluctuations work as the pairing interaction for the $s_{++}$-state, 
while the ferro-fluctuations are also favorable for the $s_{\pm}$-state.
For $J/U<(J/U)_c \equiv 0.088$,
the relation $\a^s_{\rm max}<\a^c_{\bm 0}=0.97$ is realized
and $\a^c_{\Q}$ increases towards unity.
%because of large AL term due to two-orbiton process.
In this case, orbital order occurs prior to the spin order
as increasing $U$ with $J/U$ is fixed,
since the VC (due to two-orbiton process) can efficiently enlarge
orbital susceptibilities because of large $\a^c_{\rm max}({\rm RPA})$.
This situation would be consistent with wider non-magnetic 
orthorhombic phase in Nd(Fe,Co)As and many 1111 compounds.

Since the present SC-VC method is very time-consuming,
we applied some simplifications:
We have verified in the self-inconsistent calculation that
${\rm Tr}\{ {\hat O}_\gamma{\hat X}(q){\hat O}_{\gamma'} \}$
with $\gamma\ne\gamma'$ is zero or very small,
especially at $\q={\bm 0}$ and $\Q$
for the reason of symmetry.
Since we are interested in the enhancement of $\chi^c_\gamma(\q)$
at $\q={\bm 0}$ and $\Q$ and the dominant
interferences between $\gamma=xz,yz,x^2-y^2$,
we calculated ${X}_{ll',mm'}(q)$
only for $\{(l,l'),(m,m')\} \in xz,yz,x^2-y^2$.
[$(l,l')\in \gamma$ means that $O_\gamma^{l,l'}\ne0$.]
That is, 
$\{(l,l'),(m,m')\}= \{(1,2),(3,4),(2,5)\}$, 
$\{(1,3),(2,4),(3,5)\}$, and $\{(1,5),(2,2),(3,3)\}$.

We stress that 
both $(J/U)_c$ and AF-orbital fluctuations
increase by considering following two factors:
The first one is the charge VC at each point of the three-point vertex
in Fig. \ref{fig:VC} (d), as a consequence of the Ward identity 
between ${\hat \Lambda}$ and ${\hat \chi}^{\rm irr}$.
%${\hat \Lambda}= \delta{\hat \chi}^{\rm irr}/\delta{\hat G}$.
%with respect to ${\hat \chi}^{\rm irr}(q)$.
The enhancement factor at each point is estimated as
$1+X_\gamma^c/\chi_\gamma^0=1.3\sim2.5$ for $\gamma=xz$ and $x^2-y^2$
in the present calculation near the critical point.
This effect will increase $(J/U)_c$ sensitively.
%(In LaFeAsO and BaFe$_2$As$_2$, ${\bar J}/{\bar U} \approx0.15$ is 
%predicted by the first-principle study \cite{Miyake-Arita}.)
The second factor is the $e$-ph interaction:
We introduce the quadrupole interaction in Eq. (\ref{eqn:Hquad})
%brought by the $e$-ph interaction
due to Fe-ion oscillations
 \cite{Kontani-RPA,Onari-FLEX,Saito-RPA}.
As shown in Fig. \ref{fig:chiQ} (b),
very strong AF-orbital fluctuations are obtained 
for $J/U=0.09$ and $g=0.05$; 
$\a^s_{\rm max}=\a^c_{\bm 0}=0.97$ and $\a^c_{\Q}=0.96$.
The corresponding dimensionless coupling is just
$\lambda=gN(0)\sim 0.035$ \cite{Kontani-RPA,Onari-FLEX}.
%For $g\sim0.1$ and $\beta=1$,
%$(J/U)_c$ increases to ${\bar J}/{\bar U}\approx 0.12-0.15$ 
%given by the first-principle study \cite{Miyake-Arita}.
%so we can expect the development of 
%orbital fluctuations in iron-based superconductors.
We also study the case $n=6.05$ and $g=0.065$,
and find that the relation $\a^s_{\rm max}=\a^c_{\rm max}=0.97$
is realized at $(J/U)_c=0.11$, as shown in Fig. \ref{fig:chiQ} (c).
For these reasons,
strong ferro- and AF-orbital-fluctuations would be realized 
by the cooperation of the Coulomb and weak $e$-ph interactions.

%comment
Finally, we make some comments:
The present multi-fluctuation mechanism
is not described by the dynamical-mean-field theory (DMFT),
since the irreducible VC is treated as {\it local}.
%LDA
Also, the local density approximation (LDA),
in which the VC is neglected, does not reproduce
the nonmagnetic orthorhombic phase.
Although Yanagi {\it et al}. studied 
%unphysical
%unrealistic
$U'>U$ model \cite{Yanagi} based on the RPA, 
that was first studied in Ref. \cite{Takimoto},
$\chi^c_{3z^2-r^2}({\bm 0})$ develops while 
$\chi^c_{x^2-y^2}({\bm 0})$ remains small,
inconsistently with the structure transition.
Our important future issue is to include 
the electron self-energy correction into the SC-VC method,
which is important to discuss the filling and $T$-dependences of 
orbital and spin fluctuations,
and to obtain more reliable $(J/U)_c$.

% summary
In summary, we developed the SC-VC method,
and obtained the
Coulomb-interaction-driven nematic and AF-orbital fluctuations 
due to the multimode (orbitons+magnons) interference effect
\cite{Kontani-Soft} that is overlooked in the RPA.
For $J/U\lesssim (J/U)_c$,
the structure transition ($\a^c_{\bm 0}\approx1$)
occurs prior to the magnetic transition ($\a^s_\Q\approx1$),
consistently with experiments.
%Also, impurity-induced crossover from $s_\pm$-state to $s_{++}$-state
%might be realized for $\a^s_{\rm max}\sim \a^c_{\rm max}$
When $\a^s_{\rm max}\sim \a^c_{\rm max}$, 
both $s_{++}$- and $s_\pm$-states could be realized,
depending on model parameters like the impurity concentration
\cite{Kontani-RPA,Hirschfeld}.
%Therefore, the RPA underestimates ${\hat \chi}^c(\q)$ 
%in multiorbital systems.
In a sense of the renormalization group scheme,
the quadrupole interaction in Eq. (\ref{eqn:Hquad}) is induced
by the Coulomb interaction beyond the RPA.
We expect that orbital-fluctuation-mediated 
superconductivity and structure transition 
are realized in many iron-based superconductors
due to the cooperation of the Coulomb and $e$-ph interactions.

\acknowledgements
This study has been supported by Grants-in-Aid for Scientific 
Research from MEXT of Japan, and by JST, TRIP.
Part of numerical calculations were
performed on the Yukawa Institute Computer Facility.

%%%%%%%%%%%%%%%%%%%%%%%%
%references
%%%%%%%%%%%%%%%%%%%%%%%%


\begin{thebibliography}{99}

\bibitem{Kuroki}
%K. Kuroki, S. Onari, R. Arita, H. Usui, Y. Tanaka, H. Kontani, and H. Aoki,
K. Kuroki {\it et al}., 
Phys. Rev. Lett. {\bf 101}, 087004 (2008).

\bibitem{Mazin}
I. I. Mazin, D. J. Singh, M. D. Johannes, and M. H. Du,
%I. I. Mazin {\it et al}., 
Phys. Rev. Lett. {\bf 101}, 057003 (2008).

\bibitem{Hirschfeld}
%P. J. Hirschfeld, M. M. Korshunov, I. I. Mazin 
P. J. Hirschfeld, {\it et al.},
Rep. Prog. Phys. {\bf 74}, 124508 (2011). 

\bibitem{Chubukov}
%A. V. Chubukov, D. V. Efremov, and I. Eremin,
%A. V. Chubukov {\it et al.},
%Phys. Rev. B {\bf 78}, 134512 (2008). 
A. V. Chubukov, arXiv:1110.0052.

\bibitem{DHLee}
%Fa Wang, Hui Zhai, Ying Ran, Ashvin Vishwanath, Dung-Hai Lee 
F. Wang {\it et al}., 
Phys. Rev. Lett. {\bf 102}, 047005 (2009):
The used parameters are $U=4$, $U'=2$ and $J=0.7$eV.

\bibitem{Kontani-RPA}
H. Kontani and S. Onari, 
Phys. Rev. Lett. {\bf 104}, 157001 (2010).

\bibitem{Yanagi}
%Y. Yanagi, Y. Yamakawa, and Y. Ono, 
Y. Yanagi {\it et al.},
Phys. Rev. B {\bf 81}, 054518 (2010).

\bibitem{Sato-imp}
%M. Sato, Y. Kobayashi, S. C. Lee, H. Takahashi, E. Satomi, and Y. Miura,
M. Sato {\it et al.},
J. Phys. Soc. Jpn. {\bf 79} (2009) 014710;
%S. C. Lee, E. Satomi, Y. Kobayashi, and M. Sato,
S. C. Lee {\it et al}.,
J. Phys. Soc. Jpn. {\bf 79} (2010) 023702.

\bibitem{Nakajima}
%Y. Nakajima, T. Taen, Y. Tsuchiya, T. Tamegai, H. Kitamura, and T. Murakami,
Y. Nakajima {\it et al.},
Phys. Rev. B {\bf 82}, 220504 (2010).

\bibitem{Li}
%J. Li, Y. Guo, S. Zhang, S. Yu, Y. Tsujimoto, H.Kontani, K. Yamaura, 
%and E. Takayama-Muromachi,
J. Li {\it et al.},
Phys. Rev. B {\bf 84}, 020513(R) (2011);
J. Li {\it et al.},
Phys. Rev. B {\bf 85}, 214509 (2012).

\bibitem{Paglione}
%K. Kirshenbaum, S. R. Saha, S. Ziemak, T. Drye, and  J. Paglione,
K. Kirshenbaum {\it et al}.,
arXiv:1203.5114.

\bibitem{Onari-impurity}
S. Onari and H. Kontani, 
Phys. Rev. Lett. {\bf 103} 177001 (2009).

\bibitem{Shimo-Science}
%T. Shimojima, F. Sakaguchi, K. Ishizaka, Y. Ishida, T. Kiss, M. Okawa, 
%T. Togashi, C.-T. Chen, S. Watanabe, M. Arita, K. Shimada, H. Namatame, 
%M. Taniguchi, K. Ohgushi, S. Kasahara, T. Terashima, T. Shibauchi, 
%Y. Matsuda, A. Chainani, and S. Shin,
T. Shimojima {\it et al.},
Sccience {\bf 332}, 564 (2011).

\bibitem{Saito-RPA}
%T. Saito, S. Onari, and H. Kontani,
T. Saito {\it et al.}
Phys. Rev. B {\bf 82}, 144510 (2010).

\bibitem{Onari-resonance}
S. Onari {\it et al.},
Phys. Rev. B {\bf 81}, 060504(R) (2010);
S. Onari and H. Kontani,
Phys. Rev. B {\bf 84}, 144518 (2011).

\bibitem{Fer}
%R.M. Fernandes, L. H. VanBebber, S. Bhattacharya, P. Chandra, 
%V. Keppens, D. Mandrus, M.A. McGuire, B.C. Sales, A.S. Sefat, 
%and J. Schmalian, 
R.M. Fernandes {\it et al.},
Phys. Rev. Lett. {\bf 105}, 157003 (2010). 

\bibitem{Goto}
%T. Goto, R. Kurihara, K. Araki, K. Mitsumoto, M. Akatsu, Y. Nemoto, 
%S. Tatematsu, and M. Sato, 
T. Goto {\it et al.},
J. Phys. Soc. Jpn. {\bf 80}, 073702 (2011).

\bibitem{Yoshizawa}
%M. Yoshizawa, R. Kamiya, R. Onodera, Y. Nakanishi, K. Kihou, H. Eisaki, and C. H. Lee, arXiv:1008.1479.
M. Yoshizawa {\it et al.}, 
Phys. Soc. Jpn. {\bf 81}, 024604 (2012).

\bibitem{neutron-phonon}
%J. L. Niedziela, D. Parshall, K. A. Lokshin, A. S. Sefat, A. Alatas, T. Egami
J.L. Niedziela {\it et al.},
Phys. Rev. B {\bf 84}, 224305 (2011).

\bibitem{ARPES}
%M. Yi, D. H. Lu, J.-H. Chu, J. G. Analytis, A. P. Sorini, A. F. Kemper, S.-K. Mo,
%R. G. Moore, M. Hashimoto, W. S. Lee, Z. Hussain, T. P. Devereaux, I. R. Fisher, Z.-X. Shen,
M. Yi {\it et al.},
PNAS {\bf 108} 6878.

\bibitem{Xray}
%Y. K. Kim, W. S. Jung, G. R. Han, K.-Y. Choi, K.-H. Kim, C.-C. Chen, T. P. Devereaux, A. Chainani, J. Miyawaki, Y. Takata, Y. Tanaka, M. Ouara, S. Shin, A. P. Singh, J.-Y. Kim, C. Kim 
Y. K. Kim {\it et al.},
arXiv:1112.2243.

\bibitem{Fisher}
%I. R. Fisher, L. Degiorgi, Z. X. Shen 
I. R. Fisher {\it et al.},
Rep. Prog. Phys. {\bf 74} 124506 (2011).

\bibitem{Matsuda}
%S. Kasahara, H.J. Shi, K. Hashimoto, S. Tonegawa, Y. Mizukami, T. Shibauchi, K. Sugimoto, T. Fukuda, T. Terashima, Andriy H. Nevidomskyy, Y. Matsuda
S. Kasahara {\it et al}.,
Nature {\bf 486}, 382 (2012).

\bibitem{Inoue-Nematic}
%Yoshio Inoue, Youichi Yamakawa, Hiroshi Kontani 
Y. Inoue {\it et al.},
Phys. Rev. B {\bf 85}, 224506 (2012).

\bibitem{PP}
%K. Kubo and P. Thalmeier, 
%J. Phys. Soc. Jpn. 78, 083704 (2009);
%C.-C. Lee, W.-G. Yin, and W. Ku, 
C.-C. Lee {\it et al.},
Phys. Rev. Lett. {\bf 103}, 267001 (2009);
%K. Sugimoto, E. Kaneshita, and T. Tohyama, 
%Weicheng Lv, Frank Kruger, Philip Phillips 
W. Lv {\it et al.},
Phys. Rev. B {\bf 82}, 045125 (2010);
K. Sugimoto {\it et al.},
J. Phys. Soc. Jpn. {\bf 80} (2011) 033706.

%\bibitem{Kanamori}
%J. Kanamori, Prog. Theor. Phys. {\bf 30}, 275 (1963).
%A.E. Bocquet {\it et al.}, Phys. Rev. B {\bf 46}, 3771 (1992).

\bibitem{Miyake-Arita}
%Takashi Miyake, Kazuma Nakamura, Ryotaro Arita, Masatoshi Imada 
T. Miyake {\it et al.},
J. Phys. Soc. Jpn. 79, 044705 (2010).

\bibitem{Onari-FLEX}
S. Onari and H. Kontani, 
Phys. Rev. B {\bf 85}, 134507 (2012).

\bibitem{Kontani-Soft}
%H. Kontani, T. Saito, and S. Onari, 
H. Kontani {\it et al.},
Phys. Rev. B {\bf 84}, 024528 (2011).

\bibitem{Moriya}
T. Moriya, {\it Spin Fluctuations in Itinerant Electron Magnetism}
(Springer-Verlag, 1985);
A. Kawabata: J. Phys. F {\bf 4} (1974) 1477.

\bibitem{Bickers}
N. E. Bickers and S. R. White
Phys. Rev. B {\bf 43}, 8044 (1991);
%Kunihisa Morita, Hideaki Maebashi and Kazumasa Miyake 
K. Morita {\it et al.},
J. Phy. Soc. Jpn. {\bf 72}, 3164 (2003).

\bibitem{Takimoto}
 T. Takimoto {\it et al.}, 
 J. Phys.: Condens. Matter {\bf 14}, L369 (2002). 

\end{thebibliography}
\end{document}